\documentclass[twocolumn,showpacs,preprintnumbers,amssymb,amsmath,superscriptaddress,letterpaper]{revtex4}[10pt]

\usepackage{graphicx}
\usepackage{dcolumn}
\usepackage{bm}
\usepackage{pstricks}
\usepackage{color}
\usepackage{epsfig}

\newcommand{\be}{\begin{equation}}
\newcommand{\ee}{\end{equation}}
\newcommand{\bea}{\begin{eqnarray}}
\newcommand{\eea}{\end{eqnarray}}

\def\draft{
}

\begin{document}
\draft

\title{High Energy Positrons From Annihilating Dark Matter}

\author{Ilias Cholis}
\affiliation{Center for Cosmology and Particle Physics, Department of Physics, New York University, 
New York, NY 10003}

\author{Lisa Goodenough}
\affiliation{Center for Cosmology and Particle Physics, Department of Physics, New York University, 
New York, NY 10003}

\author{Dan Hooper}
\affiliation{Theoretical Astrophysics Group, Fermi National Accelerator Laboratory, Batavia, IL  60510}
\affiliation{Department of Astronomy and Astrophysics, University of Chicago, Chicago, IL 60637}

\author{Melanie Simet}
\affiliation{Department of Astronomy and Astrophysics, University of Chicago, Chicago, IL 60637}
\affiliation{Theoretical Astrophysics Group, Fermi National Accelerator Laboratory, Batavia, IL  60510}

\author{Neal Weiner}
\affiliation{Center for Cosmology and Particle Physics, Department of Physics, New York University, 
New York, NY 10003}

\date{\today}

\begin{abstract}
Recent preliminary results from the PAMELA experiment indicate the presence of an excess of cosmic ray positrons above 10 GeV. In this letter, we consider possibility that this signal is the result of dark matter annihilations taking place in the halo of the Milky Way. Rather than focusing on a specific particle physics model, we take a phenomenological approach and consider a variety of masses and two-body annihilation modes, including $W^+W^-$, $Z^0Z^0$, $b \bar b$, $\tau^+ \tau^-$, $\mu^+ \mu^-$, and $e^+e^-$. We also consider a range of diffusion parameters consistent with current cosmic ray data. We find that a significant upturn in the positron fraction above 10 GeV is compatible with a wide range of dark matter annihilation modes, although very large annihilation cross sections and/or boost factors arising from inhomogeneities in the local dark matter distribution are required to produce the observed intensity of the signal. We comment on constraints from gamma rays, synchrotron emission, and cosmic ray antiproton measurements.

\end{abstract}
\pacs{95.35.+d; 98.70.Sa; 96.50.S; 95.55.Vj\hfill FERMILAB-PUB-08-347-A}
\maketitle


Dark matter in the form of a thermal relic is an appealing explanation for the approximately $85\%$ of the matter density of the universe not composed of baryons. In addition to being a natural extension of the big bang cosmology, the candidates which naturally give the appropriate relic abundance have annihilation cross sections of the order of the electroweak scale, a natural scale for new particles in theoretical frameworks which provide a solution to the hierarchy problem.

If they exist, such thermal relics are expected to be annihilating in the halo today, generating potentially observable fluxes of high energy particles, including gamma rays, electrons, positrons, and antiprotons. To this end, a number of cosmic ray and gamma ray experiments~\cite{Barwick:1997ig,AMS,CAPRICE,EGRET,Hinton:2004eu,MAGIC} have considered the search for dark matter annihilation products to be an important aspect of their science mission. Of particular interest is the satellite-based cosmic ray experiment, PAMELA~\cite{PAMELA,PAMELA2}. With its large acceptance (21.5 cm$^2$sr) and excellent particle identification, PAMELA is anticipated to measure the spectra of cosmic ray protons, antiprotons, electrons and positrons up to energies of 700 GeV, 190 GeV, 2 TeV and 270 GeV, respectively. 

Of particular interest for dark matter searches are high energy cosmic ray positrons~\cite{darkpositron0,darkpositron1,darkpositron2} and antiprotons~\cite{darkpositron2,darkantiproton1,darkantiproton2}. The spectra of such particles are generally expected to be dominated by the products of high energy cosmic ray interactions with the interstellar medium. In contrast, the spectra of protons and electrons are dominated by particles produced in astrophysical accelerators, e.g., supernovae. As a consequence, the ratios $e^+/(e^++e^-)$ and $\bar p/p$ are, in the absence of primary sources of cosmic ray antimatter (such as nearby pulsars~\cite{pulsars,pulsars2,pulsars3}), expected to fall at high energies. A signal of an upturn in these ratios would constitute strong evidence for a new primary source, such as dark matter annihilations.

Early results of the PAMELA experiment~\cite{PAMELATALKS} show a dramatic upturn in the positron fraction from 10 to 50 GeV (consistent with earlier indications from HEAT~\cite{Barwick:1997ig} and AMS-01~\cite{AMS}), while showing no excess in the antiproton data. These data immediately invite interpretation within the context of dark matter~\cite{bergstrom,barger,cirelli,Chen:2008yi}. Such an interpretation, however, is not trivial. In particular, the observed positron spectrum is somewhat harder than the spectral shape expected from dark matter annihilations to hadronic modes. Furthermore, the amplitude of the signal is very large and potentially difficult to reconcile with the expectations of a thermal relic. 



To calculate the cosmic ray spectra resulting from dark matter annihilations, we use the publicly available code, GALPROP~\cite{galprop}.  For a given choice of the diffusion coefficient, boundary conditions, energy loss rate, and cosmic ray injection spectrum and distribution, this code solves numerically the steady-state diffusion-loss equation\footnote{Within GALPROP, a number of additional elements are included, such as momentum space diffusion. The most relevant physical terms at high energies are included in eq \ref{dif}, however.}:
\begin{eqnarray}
0 = \vec{\bigtriangledown} \cdot \bigg[K(E)  \vec{\bigtriangledown} \frac{dn}{dE} \bigg]
+ \frac{\partial}{\partial E} \bigg[b(E,\vec{x})\frac{dn}{dE}  \bigg] + Q(E,\vec{x}),
\label{dif}
\end{eqnarray}
where $dn/dE$ is the number density of particles per unit energy, $K(E)$ is the diffusion constant, and $b(E,\vec{x})$ is the energy loss rate. The source term, $Q(E, \vec{x})$, reflects the mass, annihilation cross section, dominant annihilation modes, and distribution of dark matter in the Galaxy.

We have adopted three sets of diffusion parameters which we have found to provide provide good fits to B/C and sub-Fe/Fe cosmic ray
data above 5 GeV, and Be$^{10}$/Be$^9$ data above 1 GeV (for a review, see Ref.~\cite{review}). Thoughout this letter, we will refer to the following as Models A, B and C:
\begin{itemize}
\item[A:] {$K(E)=5.3\cdot 10^{28}$ cm$^2$/s ($E/4 \, {\rm GeV})^{0.43}$, $L=$4 kpc}
\item[B:] {$K(E)=1.4\cdot 10^{28}$ cm$^2$/s ($E/4 \, {\rm GeV})^{0.43}$, $L=$1 kpc}
\item[C:] {$K(E)=7.3\cdot 10^{28}$ cm$^2$/s ($E/4 \, {\rm GeV})^{0.43}$, $L=$6 kpc},
\end{itemize}
where $L$ is the distance above and below the galactic plane at which charged particles freely escape the Galactic Magnetic Field. For the electron/positron energy loss rate, we adopt the GALPROP defaults which include losses due to synchrotron and Inverse Compton scattering, with a local magnetic field of $5\mu G$. For the source term, we adopt an Narvarro-Frenk-White (NFW) halo profile~\cite{NFW} with a local density of 0.35 GeV/cm$^3$.


\begin{figure*}[t]
\centering\leavevmode
\includegraphics[width=2.56in,angle=0]{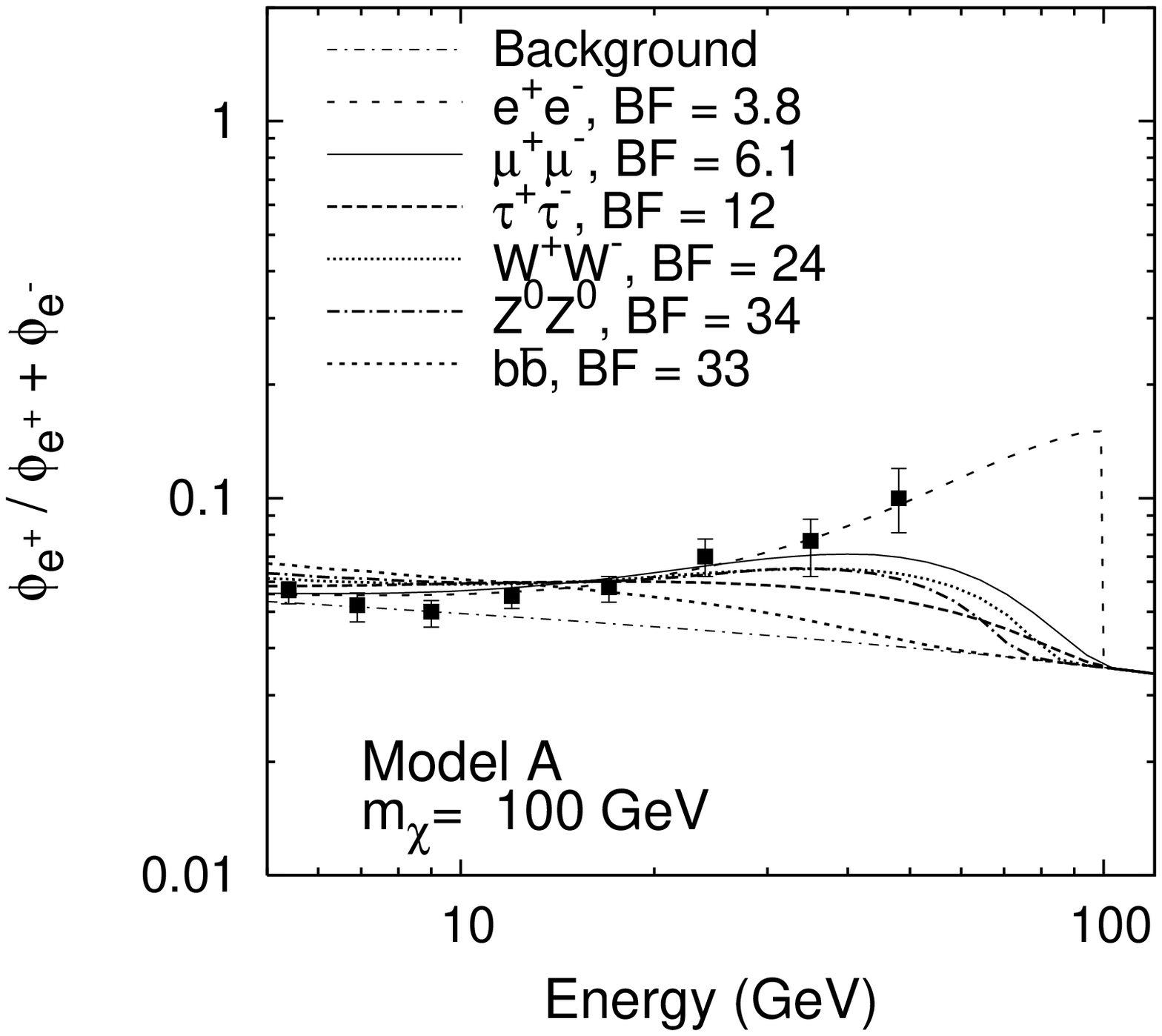}
\hspace{-1.0cm}
\includegraphics[width=2.56in,angle=0]{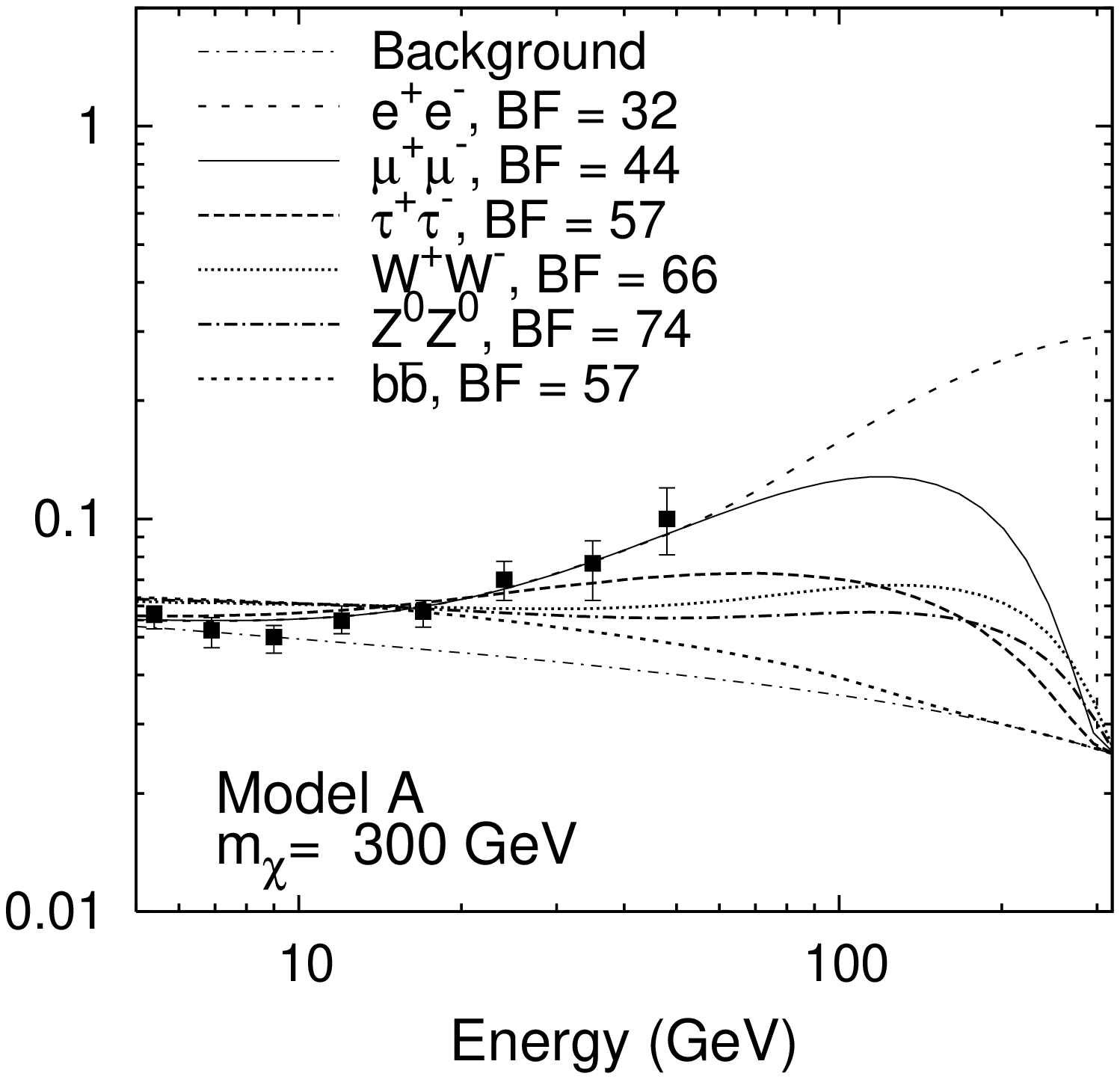}
\hspace{-1.0cm}
\includegraphics[width=2.56in,angle=0]{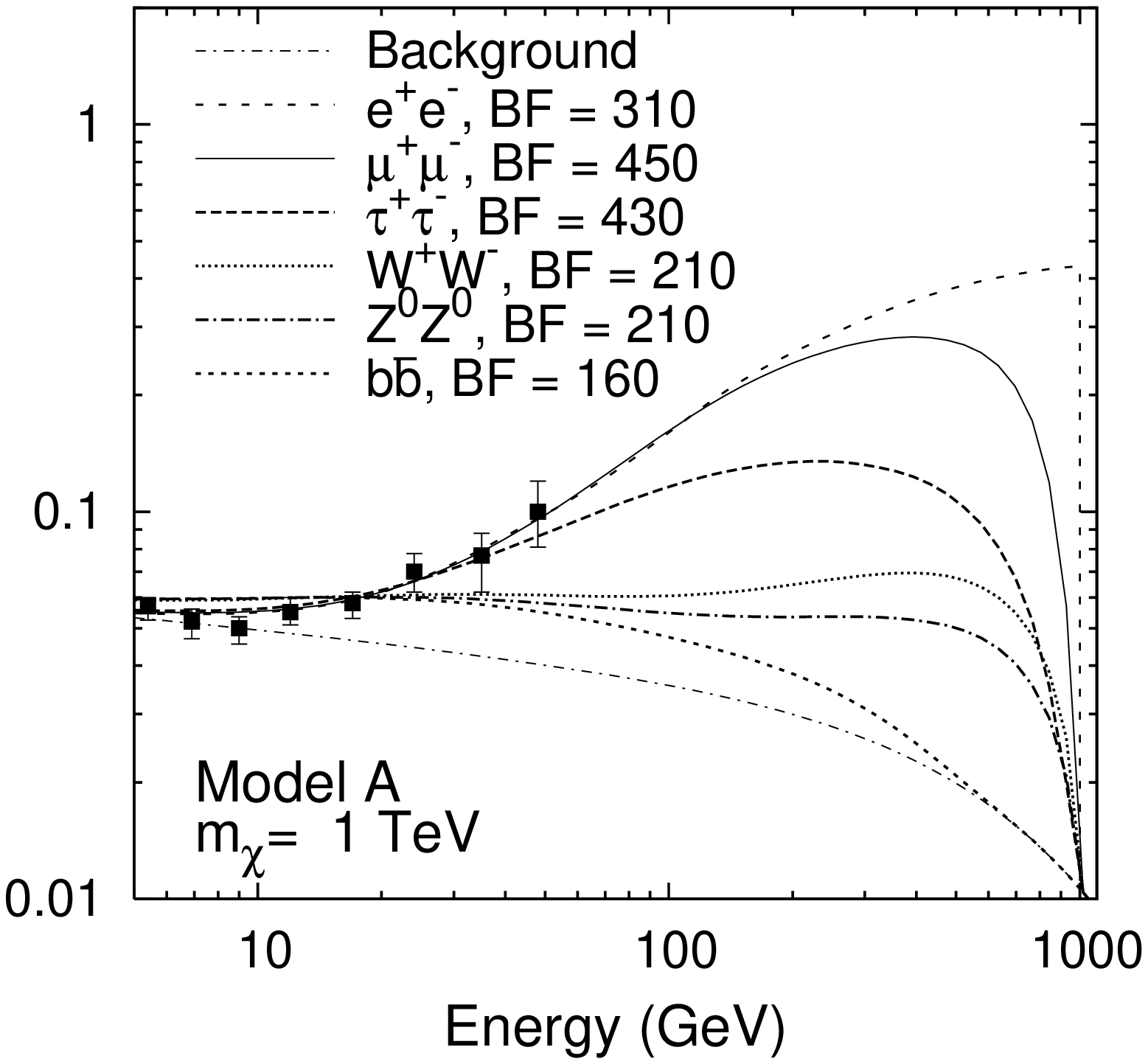}\\
\includegraphics[width= 2.56in,angle=0]{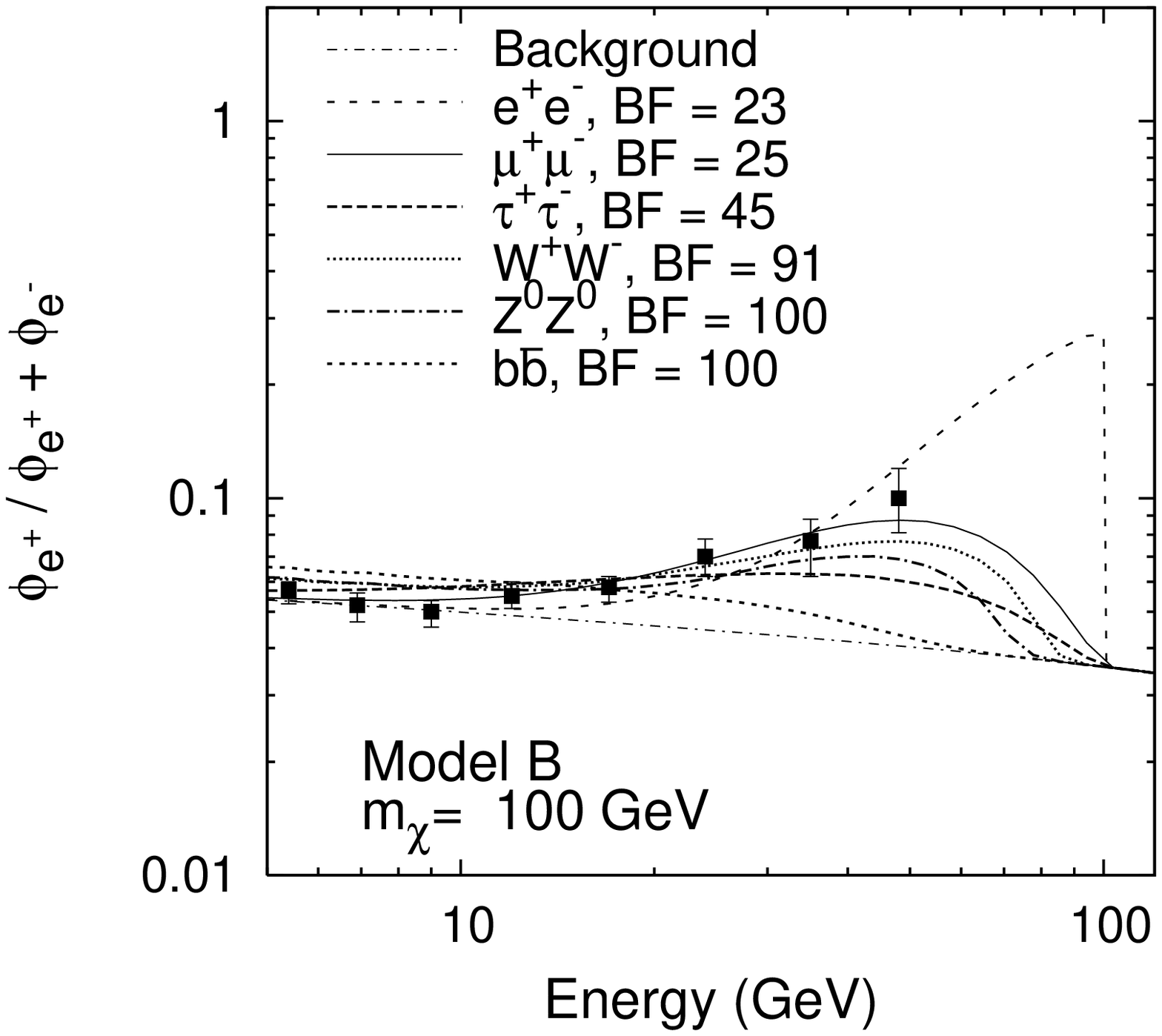}
\hspace{-1.0cm}
\includegraphics[width= 2.56in,angle=0]{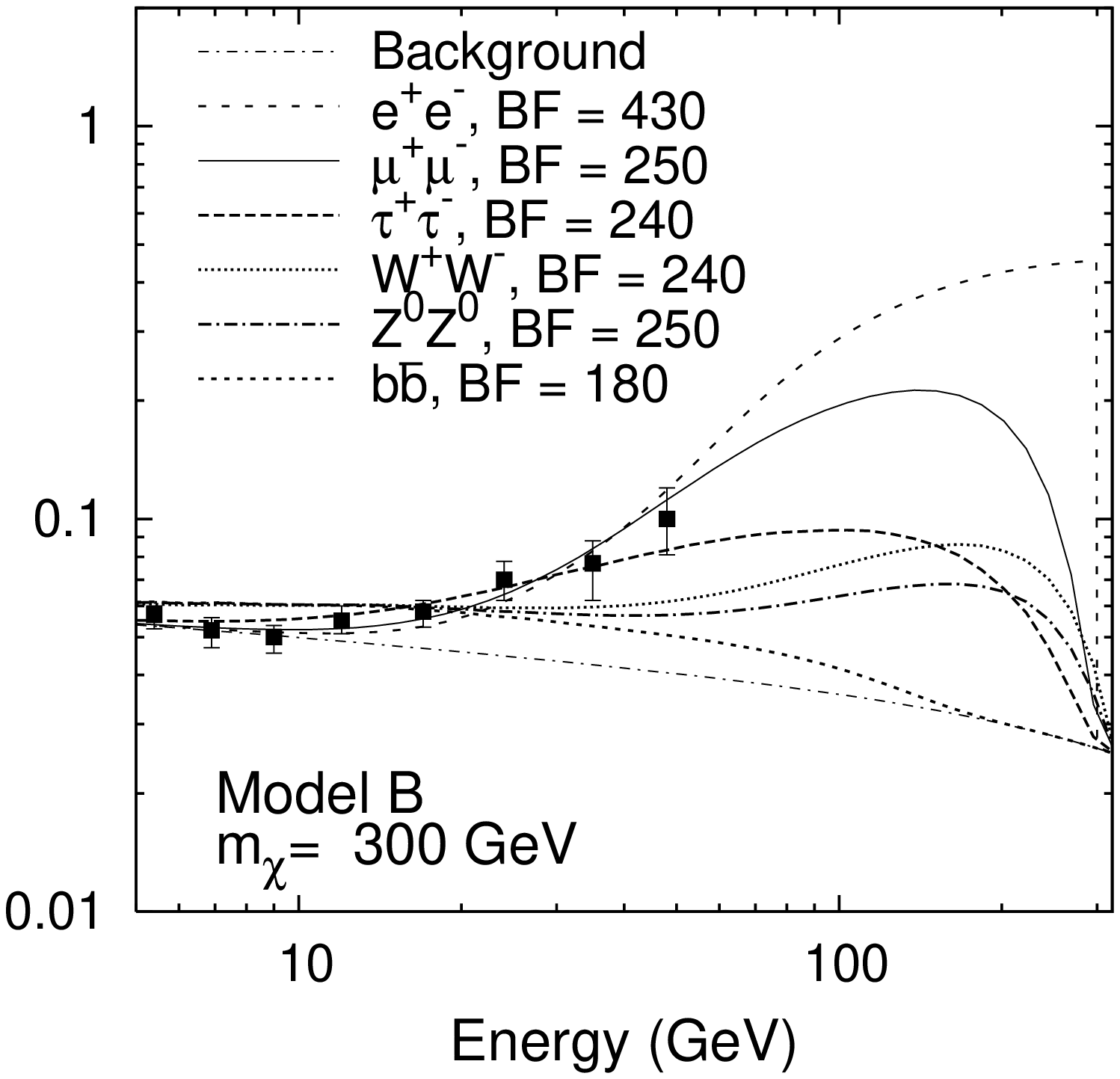}
\hspace{-1.0cm}
\includegraphics[width= 2.56in,angle=0]{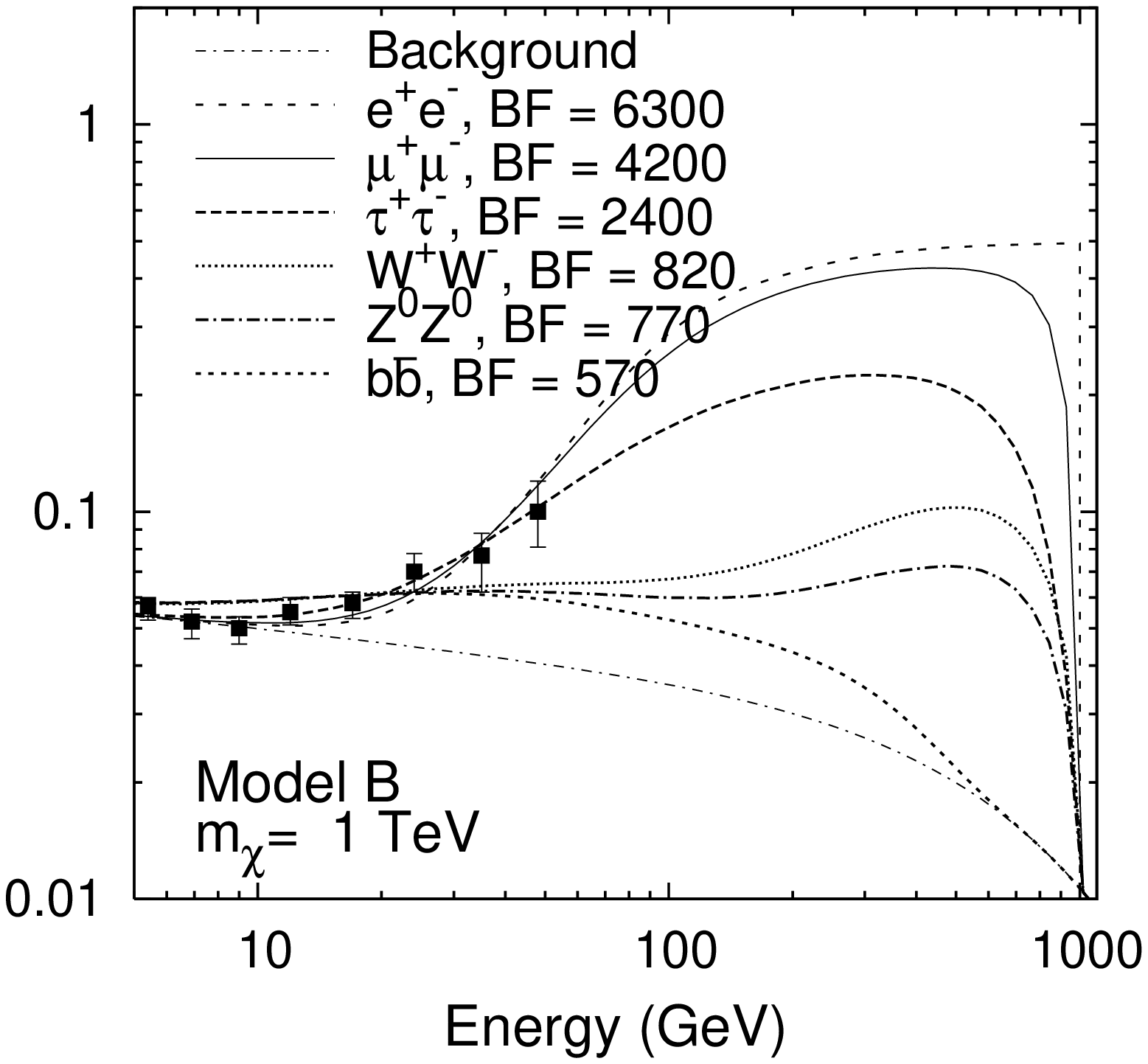}\\
\includegraphics[width= 2.56in,angle=0]{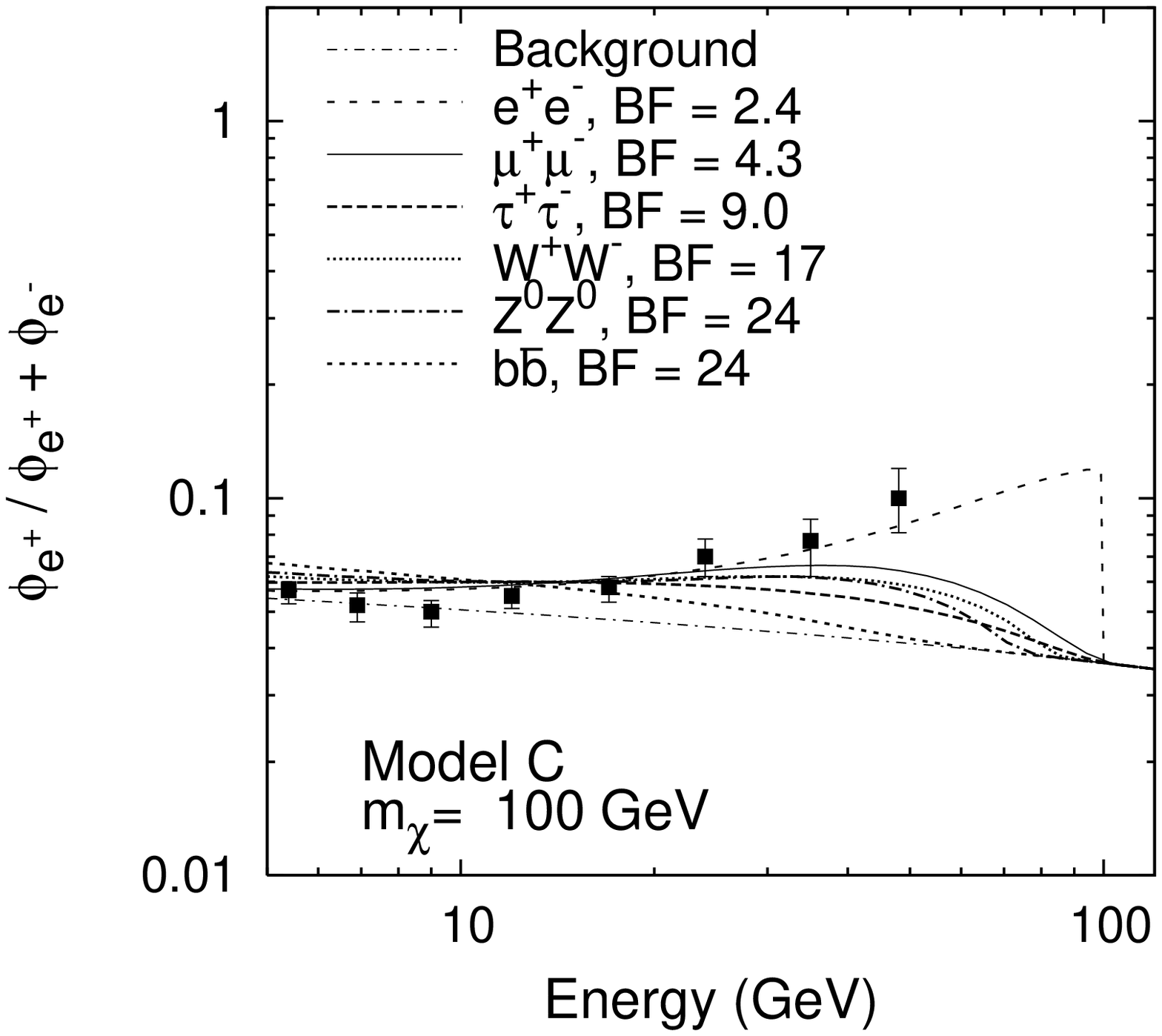}
\hspace{-1.0cm}
\includegraphics[width= 2.56in,angle=0]{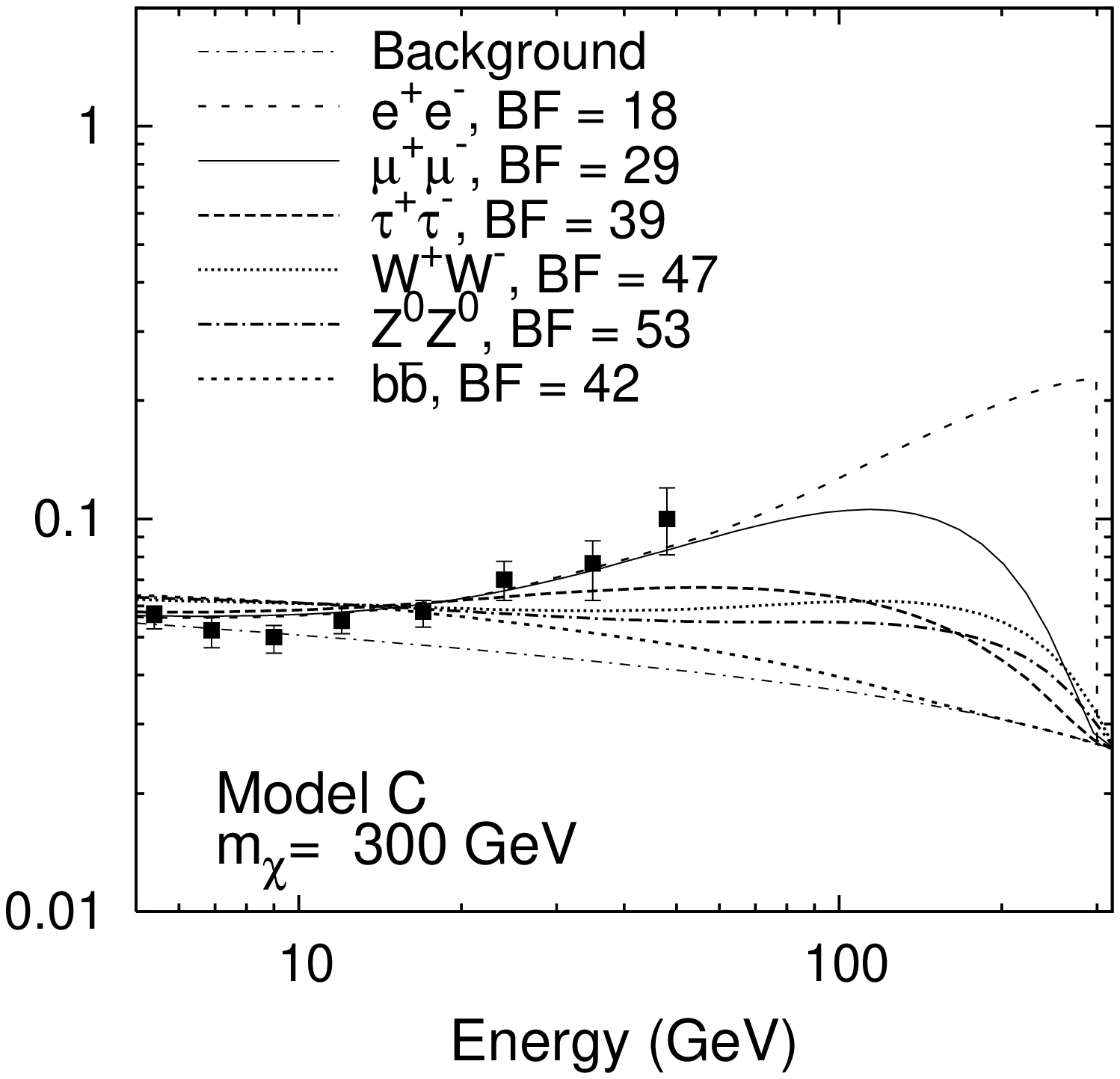}
\hspace{-1.0cm}
\includegraphics[width= 2.56in,angle=0]{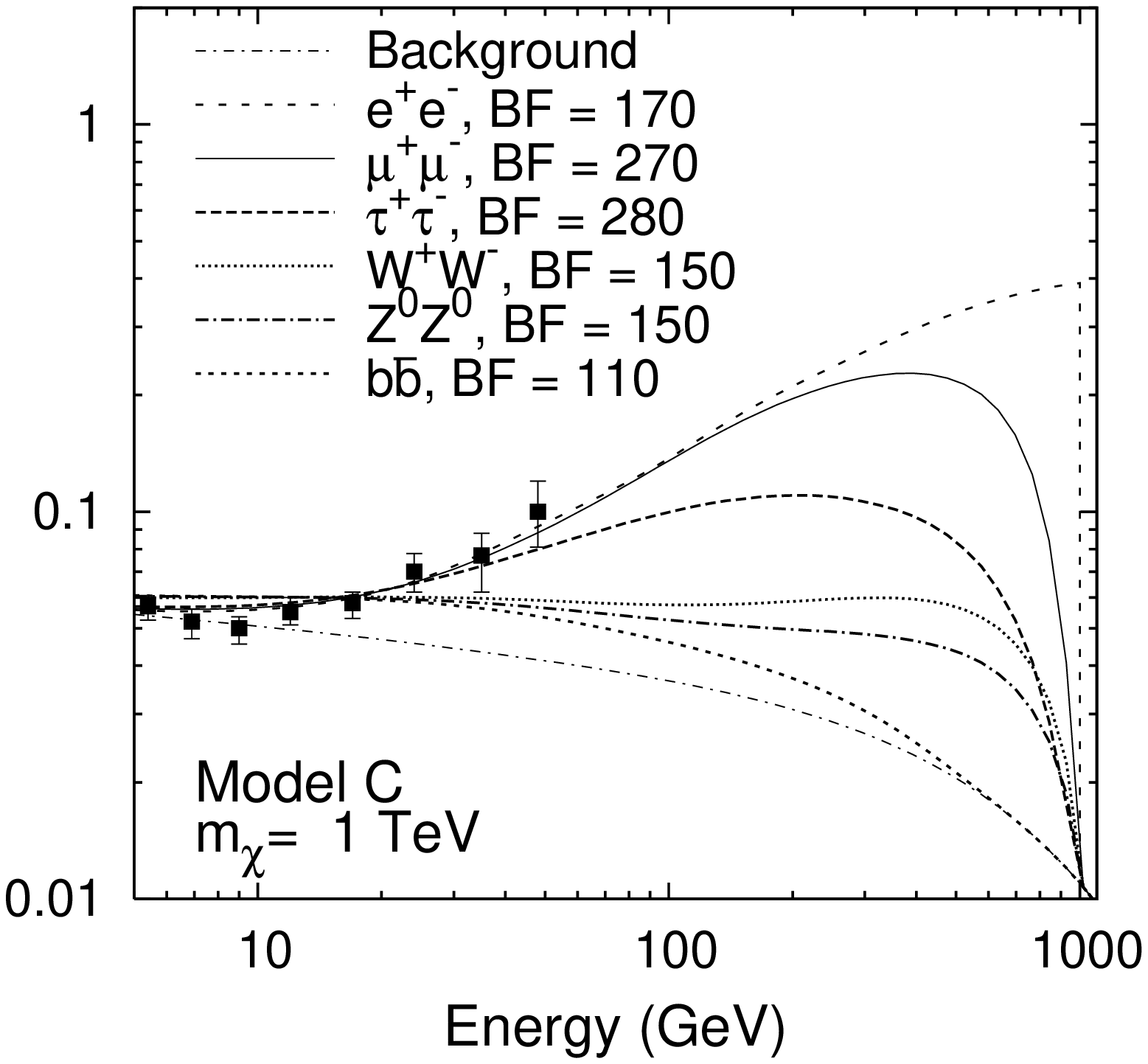}
\caption{The positron fraction as a function of energy for various dark matter masses, annihilation modes and diffusion parameters, compared to the background from secondary production alone (bottom line). In each frame, the annihilation rate was chosen to produce the best fit to the PAMELA data above 10 GeV. The required boost factor was calculated using our default values for the annihilation cross section ($\sigma v = 3 \times 10^{-26}$ cm$^3$/s) and the local dark matter density (0.35 GeV/cm$^3$).}
\label{main}
\end{figure*}


In Fig.~\ref{main}, we plot the positron fraction as a function of energy for various dark matter masses, annihilation modes, and diffusion parameters. In each case, we have normalized the dark matter annihilation rate to provide the best possible fit to the PAMELA data above 10 GeV  \cite{PAMELATALKS} (below which the effects of charge dependent solar modulation can be significant). Although the PAMELA data are still preliminary, publicly available interpretations of these preliminary data have already appeared in the literature \cite{bergstrom,barger,cirelli}, and we follow this approach. The normalization of the annihilation rate is proportional to the annihilation cross section, the square of the local dark matter density, and a quantity known as the boost factor, which parameterizes the effects of clumps and other inhomogeneities in the local dark matter distribution relative to a smooth NFW profile. In each frame of Fig.~\ref{main}, we adopt a default value for the dark matter annihilation cross section ($\sigma v = 3 \times 10^{-26}$ cm$^3$/s) and the local density (0.35 GeV/cm$^3$), and vary the boost factor to obtain the required normalization.

\begin{table}[t]
\begin{tabular}{lc|cc|cc|cc}
\hline
&& \multicolumn{2}{|c|}{Model A} & \multicolumn{2}{|c|}{Model B} &\multicolumn{2}{c}{Model C}  \\
\hline
Mass  & Mode & $\chi^2/df$  & BF &  $\chi^2/df$  & BF  &  $\chi^2/df$  & BF   \\
\hline \hline
100 & $e^+ e^-$ & 0.152  & 3.8  & 1.459  & 23   & 0.555  & 2.4  \\
100 & $\mu^+ \mu^-$ & 1.028  & 6.1  & 0.175  & 25   & 1.577  & 4.3  \\
100 & $\tau^+ \tau^-$ & 2.893  & 12  & 2.019  & 45 & 3.224  & 9.0  \\
100 & $W^+ W^-$ & 1.758  & 24 & 0.728  & 91  & 2.259  & 17   \\
100 & $ZZ$ & 1.921  & 34 & 1.139  & 100   & 2.413  & 24  \\
100 & $b \bar{b}$ & 5.154  & 33  & 4.692  & 100   & 5.107  & 24   \\
\hline
300 & $e^+ e^-$ & 0.182  & 32 & 1.132  & 430  & 0.439  & 18  \\
300 & $\mu^+ \mu^-$ & 0.186  & 44 & 0.475  & 250  & 0.532  & 29   \\
300 & $\tau^+ \tau^-$ & 1.131  & 57  & 0.387  & 240  & 1.586  & 39   \\
300 & $W^+ W^-$ & 2.598  & 66 & 2.483  & 240  & 2.781  & 47   \\
300 & $ZZ$ & 3.126  & 74 & 2.993  & 250   & 3.256  & 53  \\
300 & $b \bar{b}$ & 4.133  & 57 & 3.735  & 180  & 4.216  & 42   \\
\hline
1000 & $e^+ e^-$ & 0.106  & 310  & 1.533  & 6300  & 0.210  & 170  \\
1000 & $\mu^+ \mu^-$ & 0.128  & 450 & 0.902  & 4200  & 0.339  & 270  \\
1000 & $\tau^+ \tau^-$ & 0.333  & 430  & 0.118  & 2400   & 0.693  & 280  \\
1000 & $W^+ W^-$ & 2.243  & 210  & 1.757  & 820  & 2.515  & 150  \\
1000 & $ZZ$ & 2.552  & 210  & 2.055  & 770   & 2.809  & 150   \\
1000 & $b \bar{b}$ & 2.877  & 160 & 2.270  & 570  & 3.141  & 110  \\
\hline \hline
\end{tabular}
\caption{The quality of the spectral fit ($\chi^2$ per degree of freedom) 
and the boost factors required for various dark matter masses (in GeV), annihilation modes, and diffusion parameters to produce the PAMELA positron excess. The column BF contains the boost factors 
required assuming a local dark matter density of $\rho=0.35$. As stated in the text, the $\chi^2/df$ should be interpreted as a qualitative distinction between the scenarios, as the data are still preliminary and errors only statistical.}
\label{param-table}
\end{table}

We find that dark matter annihilations to leptons (in particular to $e^+ e^-$ and $\mu^+ \mu^-$) are naturally able to provide a good fit to the spectral shape observed by PAMELA. Annihilations to gauge bosons or quarks, however, tend to produce too soft a spectrum. This can be ameliorated if the diffusion boundary is small enough to limit the contribution from more distant annihilations (diffusion model B). In Table I, we give a qualitative measure of the quality of the fit to the PAMELA spectrum (the $\chi^2$ per degree of freedom for the data points above 10 GeV) for each case. As the data do not include systematic errors, and as they are presently highly preliminary, these numbers should not be taken as any rigorous measure of acceptable fits, but simply a means to distinguish the different scenarios. We also show in Table I the boost factor required to normalize each case to the PAMELA spectrum, assuming a local dark matter density of 0.35 GeV/cm$^3$.

The annihilation rate required to generate the flux of positrons observed by PAMELA is quite large and thus can be constrained by cosmic ray antiproton, synchrotron~\cite{Hooper:2008zg,kane}, and gamma ray measurements, especially those of EGRET~\cite{glast}. In particular, if the large annihilation rate is the result of the dark matter possessing a large annihilation cross section (such as a non-thermally produced wino in an anomaly mediated supersymmetric scenario, for example~\cite{Moroi:1999zb,kane}), then constraints on diffuse gamma rays from EGRET would likely be exceeded by a factor of two or more if those annihilations proceed to heavy quarks or gauge bosons. This conclusion can be evaded, however, if annihilations proceed largely to leptons. If the halo profile is cuspy, annihilations to gauge bosons are tightly constrained by synchrotron radiation limits in the center of the galaxy \cite{Hooper:2008zg,kane}. Both of these constraints can be evaded if the local dark matter annihilation rate is boosted by inhomogeneities in the surrounding few kiloparsecs without boosting the annihilation rate throughout the remainder of the Galaxy. 
 
Although somewhat unlikely, it is possible that the Solar System happens to be near a large dark matter subhalo, leading to a large positron flux without the overproduction of gamma rays or antiprotons throughout the halo~\cite{Hooper:2003ad}. Recent results from the Via Lactea II simulation found that although the overall annihilation rate throughout the halo is boosted by only a small value ($\sim$1.4) on average, there is small ($\sim$1\%) chance that the local annihilation rate is enhanced by more than a factor of 10 as a result of a large nearby subhalo~\cite{Diemand:2008in}. Recent results from the Aquarius Project are more pessimistic, however \cite{Springel:2008by}. Alternatively, dark matter annihilating in a density spike surrounding a nearby black hole ($M\sim 10^2$-$10^6 M_{\odot}$) could strongly boost the local annihilation rate~\cite{Brun:2007tn}. It is also interesting to note that the spectrum of positrons from a nearby subhalo or intermediate mass black hole would appear harder than if the positrons were produced from throughout a larger volume, perhaps enabling a better fit to the PAMELA data for annihilations to $b \bar{b}$ and other non-leptonic channels~\cite{Hooper:2003ad}.

{\bf In summary}, the PAMELA excess of high energy positrons, confirming earlier excesses from HEAT and AMS-01, raises the exciting possibility that we are seeing evidence of dark matter annihilations. In this letter, we have considered a range of dark matter annihilation channels and masses and find many scenarios which provide a good fit to the data. In particular, dark matter annihilations to leptons (especially $e^+ e^-$ and $\mu^+ \mu^-$) quite easily fit the observed spectrum. Annihilations to heavy quarks or gauge bosons, in contrast, provide a poorer fit to the data. This can be improved if most of the annihilations occur locally (such as is expected if the Solar System resides near a large subhalo or if the Galactic Magnetic Field confines charged particles only to a region within 1-2 kpc of the Galactic Plane). In almost every case we have considered, very large annihilation rates are required to produce the observed signal. In particular, 100 GeV (1 TeV) dark matter particles require annihilation rates boosted by a factor of approximately $\sim$2.5 to 100 ($\sim$100 to a few thousand) relative to the rate expected for a typical thermal cross section ($\sigma v \approx 3 \times 10^{-26}$ cm$^3$/s) and a smooth halo with a local density of 0.35 GeV/cm$^3$. Such boost factors could arise from inhomogeneities in the local dark matter distribution, such as the presence of a large nearby subhalo or a density spike surrounding a nearby intermediate mass black hole. 

Such a large annihilation rate could also be generated by a non-thermally produced dark matter candidate with a large annihilation cross section, but this scenario is tightly constrained by EGRET's measurements of the diffuse gamma ray spectrum (and to a somewhat lesser extent by cosmic ray antiproton and synchrotron measurements~\cite{Hooper:2008zg,kane}). For example, although non-thermal $\sim$100 GeV wino-like neutralinos could produce the measured flux of positrons, they would also exceed the diffuse gamma ray constraint by at least a factor of two (see the discussion in Ref.~\cite{glast}.) A non-thermal dark matter candidate which annihilates largely to leptons, however, could evade such constraints.

Although one could argue that the current data disfavor dark matter candidates which annihilate largely to quarks or gauge bosons, including neutralinos (for a possible exception, see Ref.~\cite{bergstrom}), and prefer those which annihilate largely to leptons, including Kaluza-Klein dark matter in models with a universal extra dimension~\cite{Cheng:2002ej,Hooper:2004xn} or ``exciting'' dark matter (XDM)~\cite{Cholis:2008vb}, uncertainties in the diffusion model make such conclusions premature. As PAMELA data become available at higher energies, however, it will become increasingly possible to discriminate between various dark matter models. Data from GLAST~\cite{glast} will also be very useful in constraining the possibility of annihilations to non-leptonic channels.

\vskip 0.2 in
\noindent {\bf Acknowledgments}
\vskip 0.05in
\noindent We thank D.~Finkbeiner and G.~Dobler for helpful discussions. DH and MS are supported by the US Department of Energy and by NASA grant NAG5-10842. NW is supported by NSF CAREER grant PHY-0449818, and IC, LG and NW are supported by DOE OJI grant \# DE-FG02-06ER41417.


\bibliography{pamela}
\bibliographystyle{apsrev}

\end{document}